\documentclass[useAMS,usenatbib]{mn2e}

\usepackage{graphics}
\usepackage{epsf}
\usepackage{subfigure}

\def\apj{Astrophysics Journal}

\def\mnras{Monthly Notices of the Royal Astronomical Society}
\def\aap{Astronomy and Astrophysics}

\title[{\it AKARI} SMC Cepheid P-L Relation]{Period-Luminosity Relations for Small Magellanic Cloud Cepheid Based on {\it AKARI} Archival Data}

\author[Ngeow et al.]{Chow-Choong Ngeow$^{1}$\thanks{E-mail: cngeow@astro.ncu.edu.tw}, Danielle M. Citro$^{2}$ and Shashi M. Kanbur$^{2}$
\\
$^{1}$Graduate Institute of Astronomy, National Central University, Jhongli City, 32001, Taiwan\\
$^{2}$Department of Physics, SUNY Oswego, Oswego, NY 13126, USA \\
}

\begin{document}

\date{Accepted 2011 October 24. Received 2011 October 24; in original form 2011 October 3}

\pagerange{\pageref{firstpage}--\pageref{lastpage}} \pubyear{2012}

\maketitle

\label{firstpage}

\begin{abstract}

In this work we matched the {\it AKARI} archival data to the Optical Gravitational Lensing Experiment-III (OGLE-III) catalog to derive the mid-infrared period luminosity (PL) relations for Small Magellanic Cloud (SMC) Cepheids. Mismatched {\it AKARI} sources were eliminated using random-phase colors obtained from  the full $I$-band light curves from OGLE-III. It was possible to derive PL relations in the $N3$ and $N4$ bands only, although the $S7$, $S11$, $L15$, and $L24$ band data were also tested. Random-phase correction was included when deriving the PL relation in the $N3$ and $N4$ bands using the available time of observations from {\it AKARI} data. The final adopted PL relations were: $N3=-3.370\log P + 16.527$ and $N4=-3.402\log P + 16.556$. However, these PL relations may be biased due to the small number of Cepheids in the sample.

\end{abstract}


\section{Introduction}

The period-luminosity (PL, also known as the Leavitt Law) relation of Cepheid variables is an important relation for determining extra-galactic distances. Previous research has focused on the PL relation of Cepheids in the optical and near infrared $JHK$ bands, with only recent emphasis on the mid-infrared. Similar to the advantages of $JHK$ PL relations \citep[see, for example,][]{mcg83}, the mid-infrared PL relations are important because: (i) they are less susceptible to extinction; (ii) they are expected to have a smaller amplitude and therefore magnitudes can be determined using a smaller number of observations; (iii) they may not be sensitive to metallicity \citep{fre08,fre10}; and (iv) they have a smaller dispersion as compared to optical PL relations. Previous work on the mid-infrared PL relations, based on {\it Spitzer} data, has concentrated on the Large Magellanic Cloud (LMC) Cepheids \citep{fre08,nge08,mad09,nge09,sco11}, the Small Magellanic Cloud \citep[SMC,][]{nge10} and Galactic Cepheids \citep{mar10}. 

In addition to {\it Spitzer}, data taken from the Infrared Camera \citep[IRC,][]{ona07} on-board the {\it AKARI} satellite \citep{mur07} can also be used to derive the mid-infrared PL relations, providing an independent cross check to {\it Spitzer} results. This has been done for the LMC Cepheids as presented in \citet[][hereafter Paper I]{nge10a}. In this paper, we continue our investigation of the mid-infrared PL relation using {\it AKARI} data for the SMC Cepheids. Since extinction is negligible in the mid-IR bands, this paper does not take extinction into account.

\section{Sample Selection}

\begin{figure}
  \centering 
  \epsfxsize=8.0cm{\epsfbox{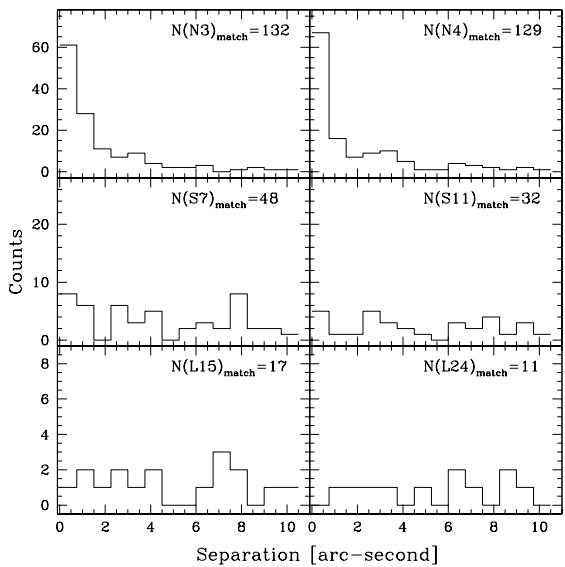}}
  \caption{Histograms of the separation for each of the bands after matching the {\it AKARI} data to the OGLE-III SMC catalog using an initial search radius of $10$ arc-second.}
  \label{separation}
\end{figure}

The {\it AKARI} data used in this work is based on the SMC bright point sources catalog presented in \citet{ita10}. Photometry in $3.2$ ($N3$, $12899$ sources), $4.1$ ($N4$, $9748$ sources), $7$ ($S7$, $1838$ sources), $11$ ($S11$, $1045$ sources), $15$ ($L15$, $479$ sources), and $24$ ($L24$, $356$ sources) $\mu \mathrm{m}$ bands provided from the {\it AKARI} catalog, along with the the time of observation, is on the IRC-Vega magnitude system as defined by \citet{tan08}. Catalog compilation and data reduction details can be found in \citet{ita10}. This catalog was matched to the OGLE-III (Optical Gravitational Lensing Experiment) SMC fundamental mode (FU) Cepheid catalog from \citet{sos10}. 

\begin{figure*}
  \vspace{0cm}
  \hbox{\hspace{-0.2cm}
    \epsfxsize=6.0cm \epsfbox{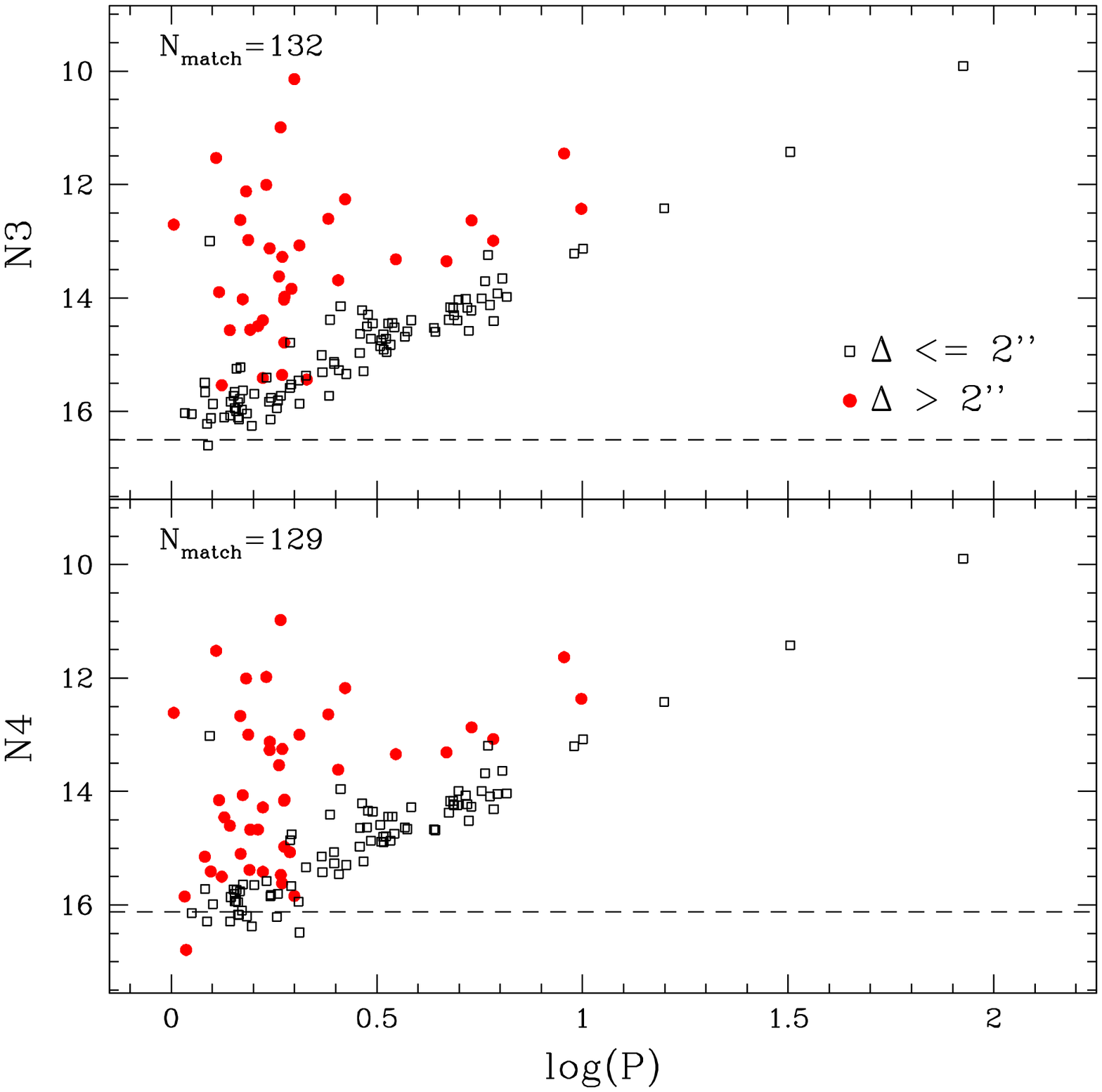}
    \epsfxsize=6.0cm \epsfbox{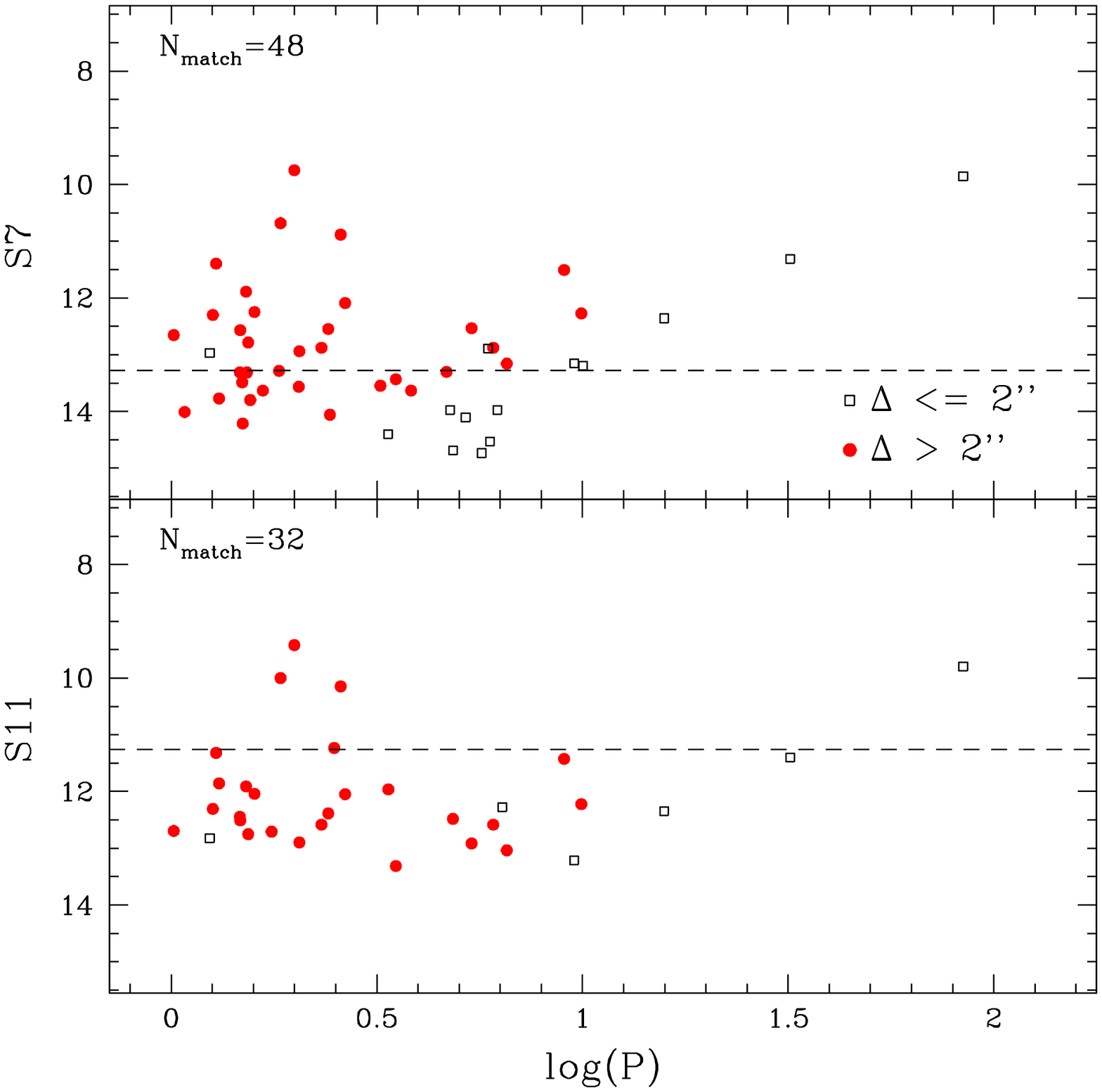}
    \epsfxsize=6.0cm \epsfbox{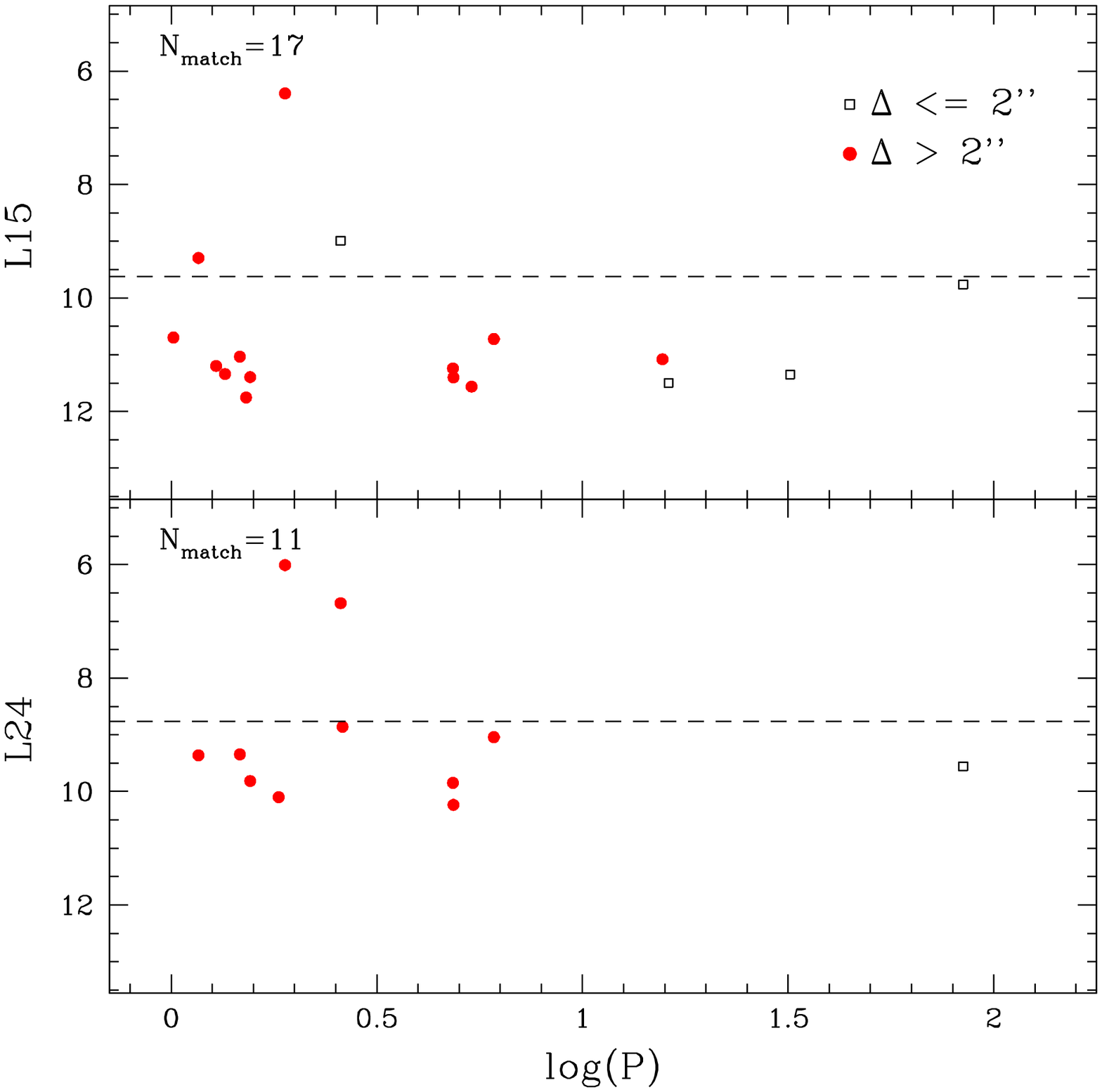}
  }
  \vspace{0cm}
  \caption{PL relations in {\it AKARI} bands based on the matched sources from \citet{ita10} catalog to the OGLE-III SMC Catalog \citep{sos10}. Initial matches were done by adopting a search radius of $10$-arcsecond. The open squares and filled circles represent the separation between catalogs ($\Delta$) that are smaller and larger than $2$-arcsecond, respectively. Dashed lines are the $10\sigma$ detection limits as listed in Table 3 of \citet{ita10}.}
  \label{rawpl}
\end{figure*}

Distributions of the separation between the matched {\it AKARI} sources and the input SMC Cepheids catalogs and the corresponding PL relations for the matched sources are presented in Figure \ref{separation} and \ref{rawpl}, respectively. \citet{ita10} did not survey the entire SMC, but instead they made $13$ individual pointings with $10'\times 10'$ field-of-view. Hence the number of matched sources is low. From these two figures, it is clear that both $N3$ and $N4$ have a definite PL relation, however, the sequence is not visible in the $S7$, $S11$, $L15$, and $L24$ bands. Furthermore, majority of the data points in the $S7$, $S11$, $L15$, and $L24$ bands could be falsely matched points for the {\it AKARI} and OGLE-III catalogs, and are approaching the detection limits in their respective bands. Therefore, only the $N3$ and $N4$ bands data were tested further for deriving the PL relations in this paper. Finally, Cepheids with $\log(P)<0.4$, where $P$ is pulsational period in days, were removed from the sample due to two reasons: (i) it has been well known that SMC Cepheids with $\log (P)<0.4$ follow a different PL relations \citep[see][and reference therein]{bau99,nge10}; and (ii) these short period Cepheids may be affected by incompleteness limits as shown in left panels of Figure \ref{rawpl}.

\subsection{Cepheids Selection Using Random-Phase Colors}

As in Paper I, the available times of observation for {\it AKARI} photometry can be converted to phases of pulsation, $\phi$, for the matched sources. Together with the knowledge of full phased $I$ band light curves from OGLE-III \citep{sos10}, mis-matching of the {\it AKARI} sources can be eliminated using the random-phase color curves (defined as $I[\phi]-AKARI[\phi]$, where $I[\phi]$ is the $I$ band magnitude that is closest to $AKARI$ photometry at phase $\phi$). 

\begin{figure}
  \centering 
  \epsfxsize=8.0cm{\epsfbox{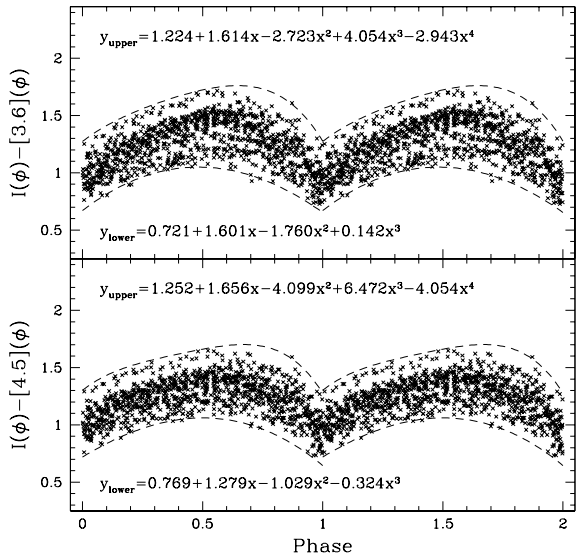}}
  \caption{Distribution of the random-phase colors for the $44$ LMC Cepheids, given in \citet{sco11}, that have full $I$ band light curves from OGLE-III LMC catalogs. Each Cepheid contains $24$ evenly-spaced data points in {\it Spitzer's} $3.6\mu\mathrm{m}$ and $4.5\mu\mathrm{m}$ bands. The dashed curves represent the boundaries of the random-phase colors distribution. The functional forms of these boundaries were given in the panels ($y$ is the random-phase colors and $x$ is the phases).}
  \label{spitzer}
\end{figure}

To guide the selection of well-matched sources from the random-phase color curves, the range of the colors as a function of phases was determined using the {\it Spitzer} light curves photometric data for LMC Cepheids available from \citet{sco11}. The authors have observed $\sim80$ LMC Cepheids with {\it Spitzer} in $3.6\mu\mathrm{m}$ and $4.5\mu\mathrm{m}$ bands, with $24$ evenly-spaced data points per light curves. For $44$ Cepheids that have the full $I$ band light curves from OGLE-III LMC Cepheid catalogs \citep{sos08}, the random-phase color curves were constructed using all $24$ {\it Spitzer} data points. Distributions of the random-phase colors as a function of phases for these Cepheids were displayed in Figure \ref{spitzer}, where the boundaries of the distribution were determined using the {\tt BOUNDFIT} code \citep{car09}.

\begin{figure*}
  \vspace{0cm}
  \hbox{\hspace{0.5cm}
    \epsfxsize=8.5cm \epsfbox{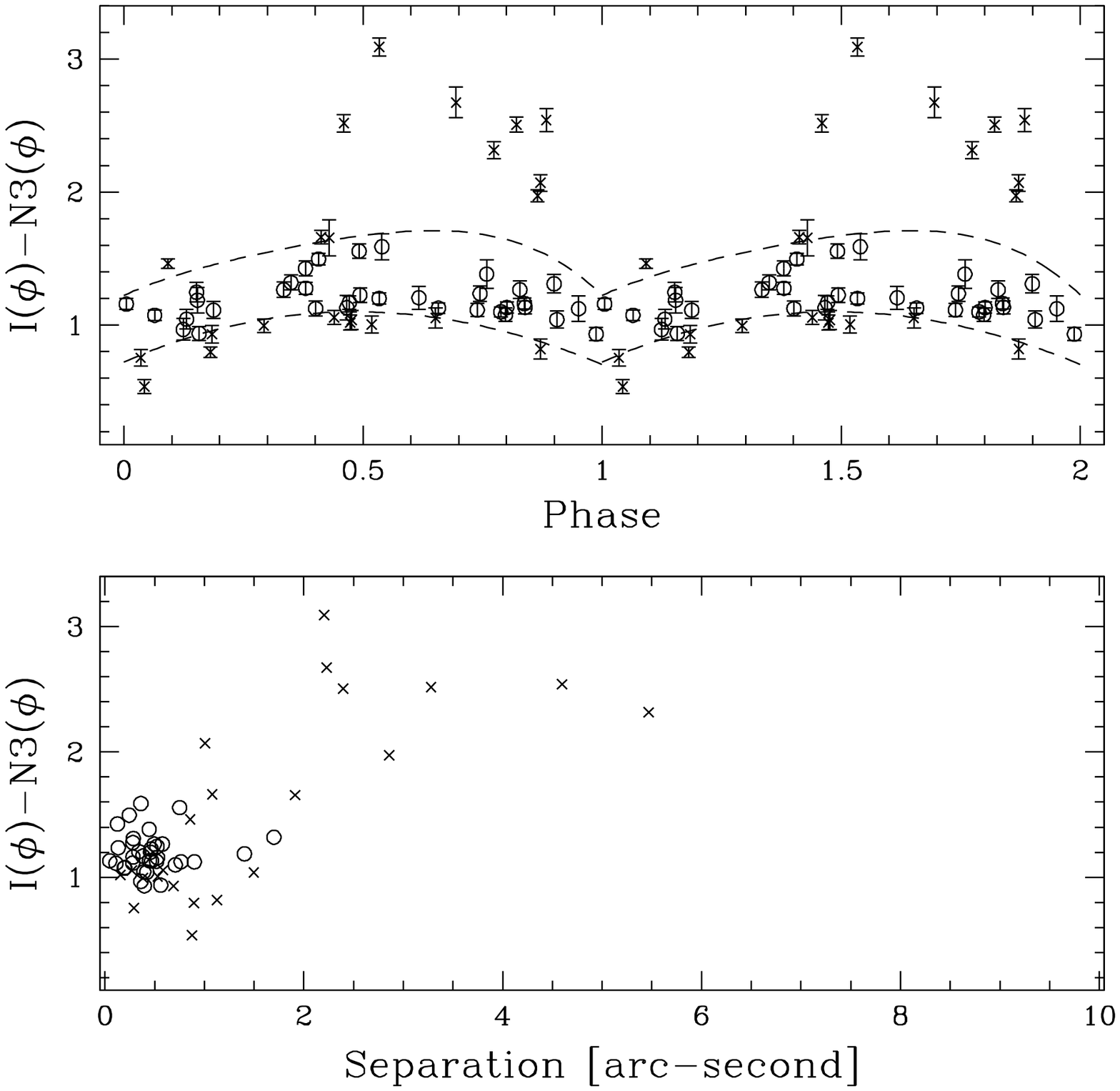}
    \epsfxsize=8.5cm \epsfbox{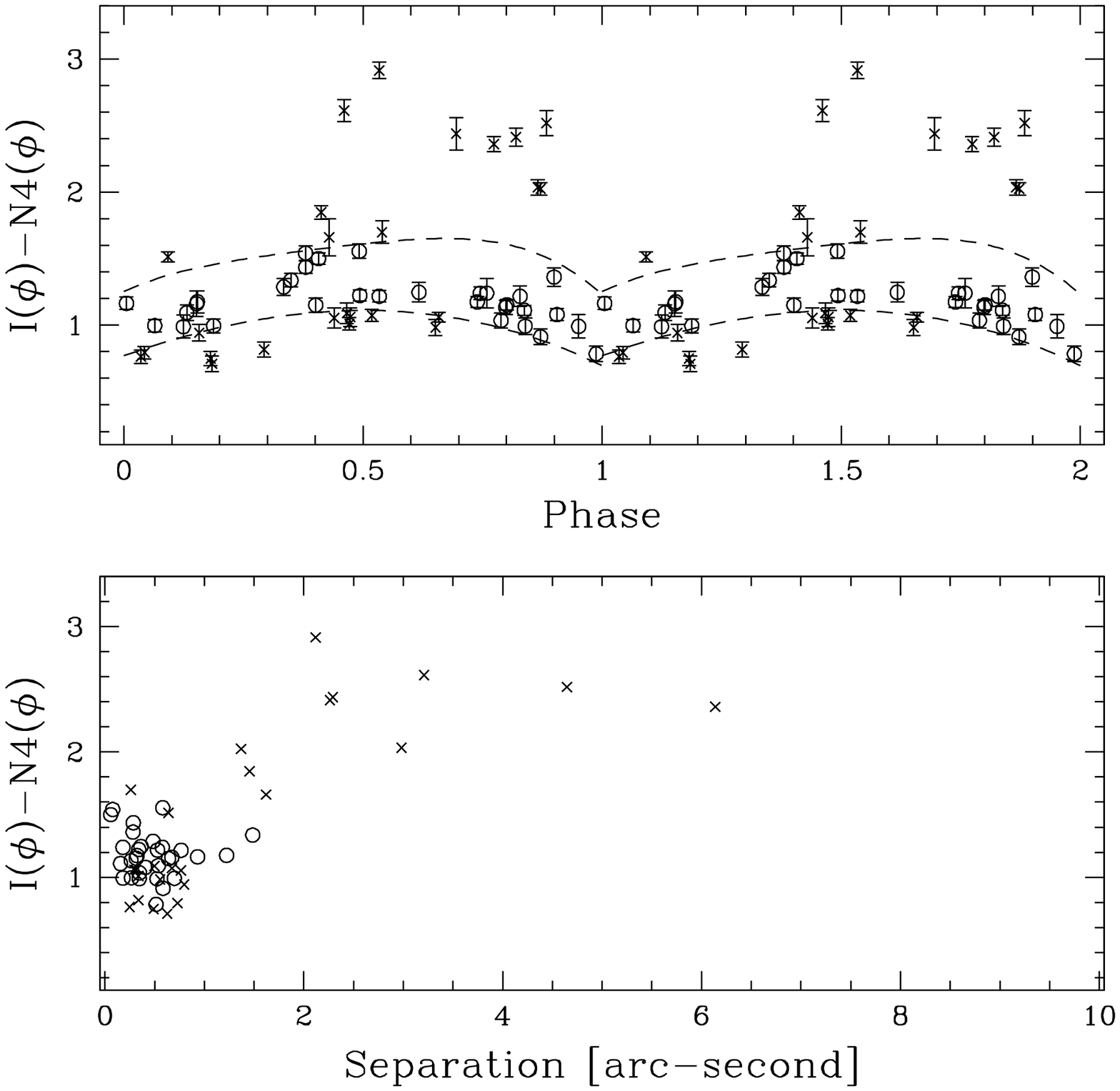}
  }
  \vspace{0cm}
  \caption{Upper panels show the random-phase colors for {\it AKARI} matched sources. Dashed curves represent the boundaries of the random-phase colors as defined in Figure \ref{spitzer}. Open circles and crosses are the matched sources located within and outside the boundaries, respectively. Lower panels show the random-phase colors as a function of separation between the input OGLE-III SMC catalog and {\it AKARI} matched sources. Error bars are omitted for clarity.}
  \label{color}
\end{figure*}

In Figure \ref{color}, the random-phase colors for matched {\it AKARI} sources were plotted as a function of phases and separations. This Figure reveals that the matched sources with $I(\phi)-AKARI(\phi)>2$ tend to have larger separations, suggesting these {\it AKARI} sources were mis-matched to the OGLE-III SMC catalog, and should be eliminated from the samples. In contrast, most of the sources with $I(\phi)-AKARI(\phi)<2$ have separations less than $2$ arc-second. Some of these sources were also located near the boundary curves defined from Figure \ref{spitzer}. If the errors in random-phase colors are considered together with the fact that {\it AKARI's} $N3$ and $N4$ bands are slightly different than the {\it Spitzer's} $3.6\mu\mathrm{m}$ and $4.5\mu\mathrm{m}$ bands, then matched sources with $I(\phi)-AKARI(\phi)<1.9$ could be used to derive the PL relations - as in the next section. This selection criterion left $49$ and $48$ matched sources in $N3$ and $N4$ band, respectively. Photometry for these sources is listed in Table \ref{tab1}. A full version of this Table will be published on-line: a part of the Table is shown here for its layout and content.

\begin{table*}
  \centering
  \caption{Photometry of the {\it AKARI} sources matched to OGLE-III SMC catalog (full version of the Table will be available on-line).}
  \label{tab1}
  \begin{tabular}{cccccc} \hline
    Cepheid & Band & Period [day] & Time of Observation & Magnitude & Magnitude Error \\
    \hline 
OGLE-SMC-CEP-1193 &	$N3$ &	3.089325 & 2454221.980058 & 14.450 & 0.097 \\
OGLE-SMC-CEP-1213 &	$N3$ &	2.995774 & 2454220.669050 & 14.498 & 0.096 \\
OGLE-SMC-CEP-1250 &	$N3$ &	4.970625 & 2454221.980058 & 14.400 & 0.044 \\
$\cdots$ & $\cdots$ & $\cdots$ & $\cdots$ & $\cdots$ & $\cdots$ \\ 
    \hline
  \end{tabular}
\end{table*}

\section{The Period-Luminosity Relations}

\begin{figure*}
  \vspace{0cm}
  \hbox{\hspace{0.5cm}
    \epsfxsize=8.5cm \epsfbox{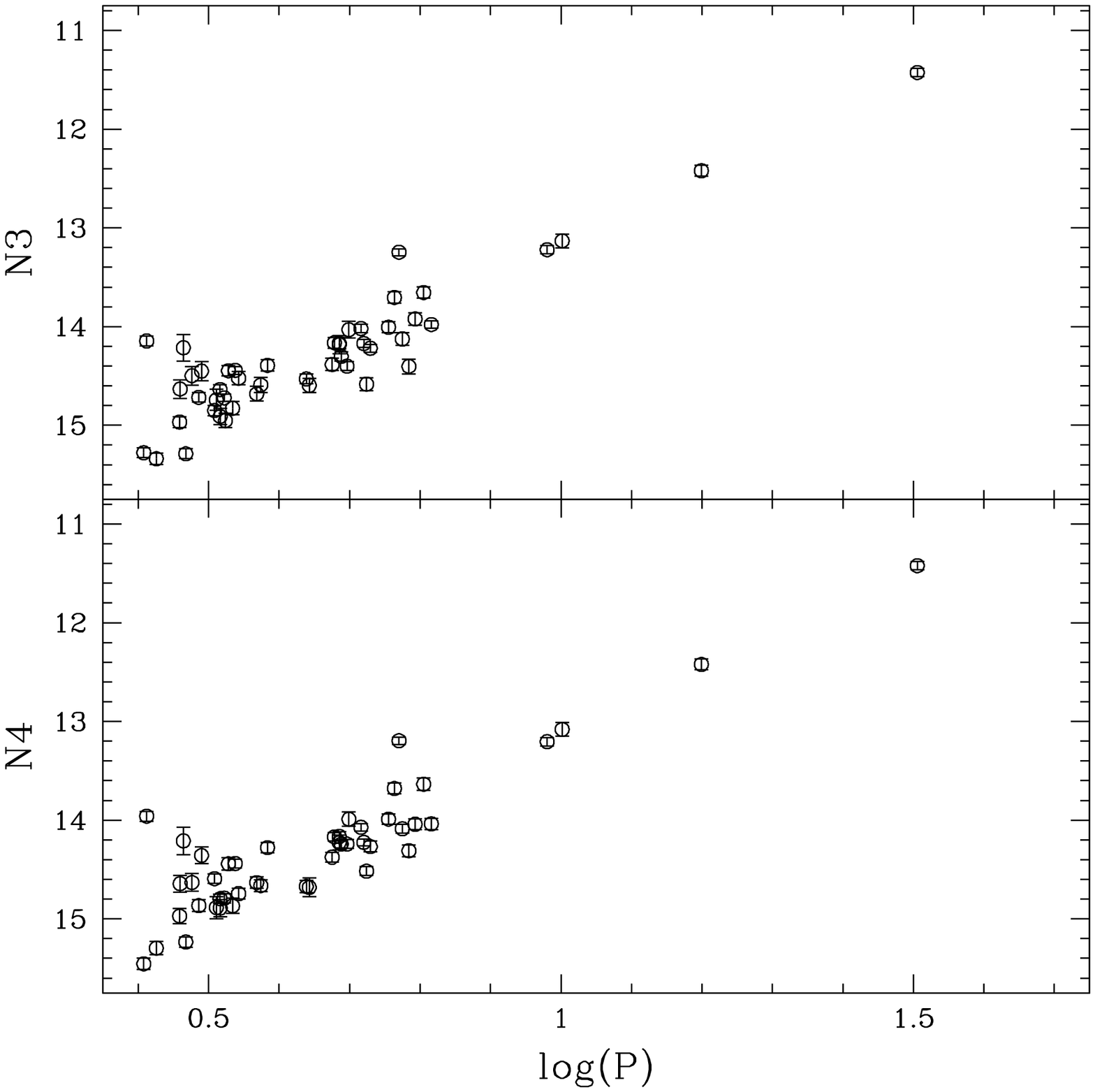}
    \epsfxsize=8.5cm \epsfbox{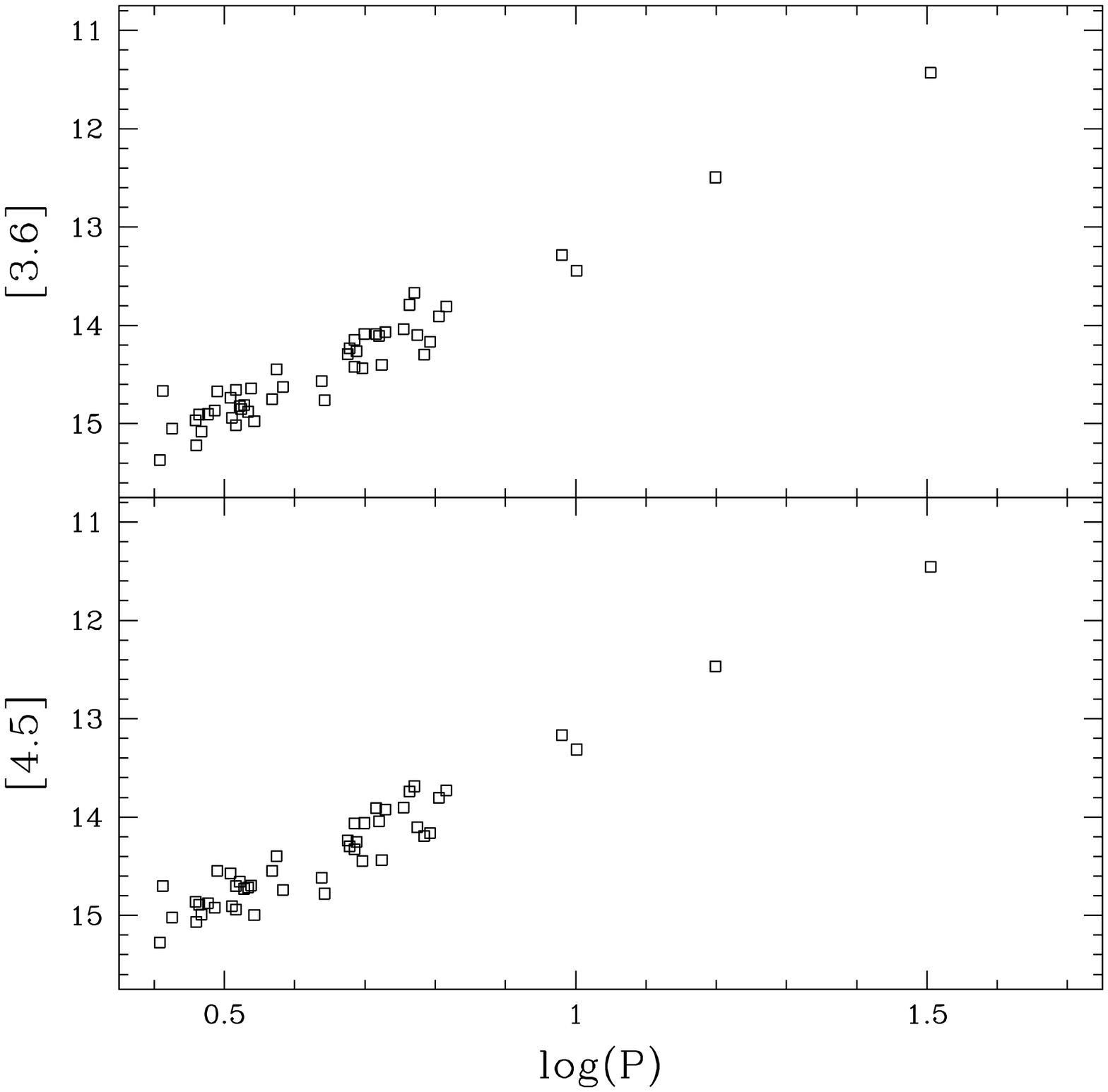}
  }
  \vspace{0cm}
  \caption{{\bf (a) Left panel:} PL relations based on the data presented in Table \ref{tab1} without random-phase corrections. {\bf (b) Right panel:} PL relations based on the {\it Spitzer} counterparts for Cepheids listed in Table \ref{tab1}. The $3.6\mu\mathrm{m}$ and $4.5\mu\mathrm{m}$ band photometry were adopted from \citet{nge10}. Error bars are omitted as they are comparable to the size of the symbols. }
  \label{plfinal}
\end{figure*}

The PL relations based on the data given in Table \ref{tab1} are presented in left panels of Figure \ref{plfinal}. The counterparts in {\it Spitzer} $3.6\mu\mathrm{m}$ and $4.5\mu\mathrm{m}$ band for the same Cepheids listed in Table \ref{tab1} were plotted in the right panels of Figure \ref{plfinal}. When comparing these PL relations, the dispersions of {\it AKARI} band PL relations are found to be larger ($\sigma\sim0.3$) than the {\it Spitzer} PL dispersions ($\sigma\sim0.2$), suggesting that few mis-matched {\it AKARI} sources may still present in the sample. Hence we applied an iterative $\sigma$-clipping algorithm to remove the outliers presented in the PL relations. The resulting random-phase PL relations are: $N3=-3.351(\pm0.113)\log P+16.518(\pm0.084)$, with $\sigma=0.210$ for $46$ Cepheids; and $N4=-3.391(\pm0.115)\log P+16.554(\pm0.086)$, with $\sigma=0.213$ for $45$ Cepheids. The $N3$ PL relations are consistent with the {\it Spitzer} $3.6\mu\mathrm{m}$ band counterpart: $[3.6]=-3.349(\pm0.091)\log P+16.570(\pm0.067)$. The $N4$ PL relation, however, is steeper than the {\it Spitzer} $4.5\mu\mathrm{m}$ band PL relation: $[4.5]=-3.326(\pm0.097)\log P+16.502(\pm0.071)$.

Knowing the time of observations for the matched {\it AKARI} sources not only permits the selection of good candidate Cepheids based on random-phase colors, but also allows the application of random-phase corrections to convert the observed random phase magnitudes to mean magnitudes. Algorithms for the random-phase corrections are the same as in Paper I, and will not be repeated here. The random-phase corrected PL relations, without including the $I$ band amplitudes in the corrections, are: $N3=-3.372(\pm0.117)\log P+16.533(\pm0.087)$ and $N4=-3.403(\pm0.119)\log P+16.562(\pm0.089)$. As in Paper I, PL relations using the random-phase corrections that include the $I$ band amplitudes term are adopted as the final PL relations:

\begin{eqnarray}
N3 & = & -3.370(\pm0.115)\log P + 16.527(\pm0.084), \nonumber \\ 
N4 & = & -3.402(\pm0.117)\log P + 16.556(\pm0.086). \nonumber
\end{eqnarray}

The slope of the SMC $N3$ PL relation is consistent with the LMC $N3$ band PL slope of $-3.246(\pm0.047)$ presented in Paper I, though it is steeper than the LMC $N3$ PL slope or the {\it Spitzer} $3.6\mu\mathrm{m}$ band PL slope for SMC Cepheids given in \citet{nge10}. For the $N4$ band PL slope, it is also found to be consistent within $\sim 2\sigma$ to the SMC $4.5\mu\mathrm{m}$ band PL slope \citep{nge10}, but the slope also tends to be steeper. The slope of the $N4$ PL relation is expected to be shallower than the $N3$ relation given the influence of the CO absorption around $\sim4\mu\mathrm{m}$ to $\sim6\mu\mathrm{m}$ \citep{mar10,sco11}. The steeper $N3$ and $N4$ band PL slopes found in this work may be caused by the small number of Cepheids present in the samples. Evidence to support this hypothesis is the slopes for a smaller subset of {\it Spitzer} $3.6\mu\mathrm{m}$ and $4.5\mu\mathrm{m}$ band PL relations ($-3.349$ and $-3.326$, respectively), as shown in right panels of Figure \ref{plfinal}, also tend to be steeper than the PL slopes from full sample as given in \citet{nge10}.

\section{Conclusion}

In this paper we derived the mid-infrared SMC PL relations based on publicly available {\it AKARI} catalog from \citet{ita10}. In contrast to the {\it Spitzer} data used in \citet{nge10}, the archival {\it AKARI} catalog includes the time of observation that permit the calculation of pulsational phases. This in turn can be used to select good Cepheid candidates based on random-phase colors and thereby derive the random-phase corrected PL relations. The steeper PL slopes found in this work may be biased due to small numbers of Cepheids in selected SMC regions, hence the application of these PL relations needs to be treated with cautious. Complete data from surveying the entire SMC using {\it AKARI} will be valuable to improve the determination of these PL relations and in the future distance scale work.

\section*{Acknowledgments}

This research is based on observations with {\it AKARI}, a JAXA project with the participation of ESA. We thank the referee for useful comments to improve this manuscript. CCN thank the funding from National Science Council (of Taiwan) under the contract NSC 98-2112-M-008-013-MY3.

\label{lastpage}

\end{document}